\documentclass[sigconf]{acmart}

\usepackage{listings}
\usepackage{xcolor} 
\usepackage{multirow}
\usepackage{makecell}

\AtBeginDocument{%
  }

\setcopyright{none}
\renewcommand\footnotetextcopyrightpermission[1]{}

\definecolor{listinggray}{gray}{0.9}
\definecolor{darkgray}{gray}{0.1}
\definecolor{lbcolor}{rgb}{0.9,0.9,0.9}
\definecolor{Darkgreen}{rgb}{0,0.4,0}
\definecolor{lightblue}{rgb}{0.4,0.4,0.8}

\lstdefinestyle{mypseudo}{
    language=,
    basicstyle=\ttfamily\small,
    commentstyle=\color{gray},            
    morecomment=[l]{//},                  
    morecomment=[s]{/*}{*/},
    numbers=left, 
    showstringspaces=false,
    tabsize=4,
    breaklines=true,
    breakatwhitespace=false,
    escapechar=!
}

\lstdefinestyle{myC}{
    language=C,
    keywordstyle=\bfseries\color{blue},
    morekeywords={u32,u64,asm},
    basicstyle=\ttfamily\small,
    commentstyle=\color{Darkgrey}, 
    numbers=left,
    showstringspaces=false,
    tabsize=4,
    breaklines=true,
    breakatwhitespace=false,
    title=\lstname,
    keywordstyle=\color[rgb]{0,0,1},
    commentstyle=\color{darkgray},
    stringstyle=\color{Darkgreen},
    morekeywords={u32,u64,asm,volatile}
}

\begin{document}

\title{Trace of the Times: Rootkit Detection through Temporal Anomalies in Kernel Activity}

\author{Max Landauer\textsuperscript{1}, Leonhard Alton\textsuperscript{1}, Martina Lindorfer\textsuperscript{2}, Florian Skopik\textsuperscript{1}, Markus Wurzenberger\textsuperscript{1}, Wolfgang Hotwagner\textsuperscript{1}}

\affiliation{\textsuperscript{1} %
	\institution{Austrian Institute of Technology}
	\city{Vienna}
	\country{Austria}
}
\email{firstname.lastname@ait.ac.at}

\affiliation{\textsuperscript{2}%
	\institution{Vienna University of Technology}
	\city{Vienna}
	\country{Austria}
}
\email{firstname.lastname@tuwien.ac.at}


\renewcommand{\shortauthors}{Landauer et al.}

\begin{abstract}
Kernel rootkits provide adversaries with permanent high-privileged access to compromised systems and are often a key element of sophisticated attack chains. At the same time, they enable stealthy operation and are thus difficult to detect. Thereby, they inject code into kernel functions to appear invisible to users, for example, by manipulating file enumerations. Existing detection approaches are insufficient, because they rely on signatures that are unable to detect novel rootkits or require domain knowledge about the rootkits to be detected. To overcome this challenge, our approach leverages the fact that runtimes of kernel functions targeted by rootkits increase when additional code is executed. The framework outlined in this paper injects probes into the kernel to measure time stamps of functions within relevant system calls, computes distributions of function execution times, and uses statistical tests to detect time shifts. The evaluation of our open-source implementation on publicly available data sets indicates high detection accuracy with an F1 score of 98.7\% across five scenarios with varying system states.
\end{abstract}



\keywords{rootkit detection, anomaly detection, kernel tracing}


\maketitle
\pagestyle{plain}

\section{Introduction} \label{intro}

Among the many types of cyber attacks that security professionals need to deal with on a regular basis, rootkits pose a particularly severe threat. Once installed, rootkits provide attackers with permanent access to compromised systems and enable to execute arbitrary commands with high (root) privileges; hence the name rootkit \cite{colbert2016cyber}. On top of that, rootkits are designed to effectively make themselves invisible to users by interfering with low-level functions of the operating system's kernel to hide their own presence \cite{bravo2011proactive, singh2017detection}. Real-world examples of cyber attacks involving rootkits can be found in many recent threat reports, for example, the 2020 CrowdStrike Global Threat Report \cite{crowdstrike2020} mentions an attack that involved a customized version of a publicly available rootkit that was used to interfere with system functions for stealthy operation. Other sources report the use of rootkits to prevent detection of cryptocurrency-mining malware \cite{remillano2019skidmap}, hiding of a malicious shared library and overwriting symbolic links \cite{leveille2024ebury}, and hiding of network connections when redirecting network traffic \cite{acsc2018hackingtools}.

Most common intrusion detection mechanisms rely on the recognition of signatures such as byte patterns that are known to correspond to certain malware \cite{bravo2011proactive}. Unfortunately, this strategy is generally not sufficient for comprehensive protection against rootkits, because no signatures are available for novel rootkits or modified versions of existing ones \cite{carreon2020statistical}. As a consequence, alternative detection methods, such as cross-view that locates discrepancies between different system levels to identify hidden objects, have been investigated in the past \cite{nadim2023kernel}.

However, in their recently published survey, St{\"u}hn et al. \cite{stuhn2024hidden} conclude that prevalent mechanisms for rootkit detection are insufficient. Their evaluation of several common intrusion detection systems demonstrates that no single solution is capable of reliably detecting various types of rootkits. In addition, they find that the detection performance of these tools heavily depends on domain knowledge about the compromised system and the deployed rootkits, which limits their applicability in real-world use-cases. Moreover, Nadim et al. \cite{nadim2023kernel} emphasize the problems of forensic approaches and express the need for an intelligent and lightweight approach that is capable of detecting novel rootkits at runtime. In this paper we therefore propose a generalized approach for real-time rootkit detection based on statistical and semi-supervised anomaly detection techniques. The main idea behind our approach is that rootkits need to inject code when interfering with the operating systems to hide their presence, which causes that the overall runtime of that modified code block increases, because additional instructions need to be executed \cite{carreon2020statistical, lu2019data}. Our detection method thus captures normal time intervals of and between executed kernel functions within system calls and recognizes significant delays with respect to these normal distributions as indicators for rootkits. Thereby, our approach goes beyond existing works that only use execution times of entire system calls, neglect multimodal features in collected data sets, and do not consider the influence of system conditions and noise \cite{luckett2016neural, dawson2018phase, brodbeck2012covert}. We emphasize that our approach does not aim to replace any existing security measures against rootkits, such as signature-based detection approaches or cross-view, but instead provides an additional line of defense that complements state-of-the-art detection systems.

There are several challenges in designing a reliable detection system based on shifts in distributions of function timings. On the one hand, there is a vast number of potentially relevant functions in the kernel that could be considered for analysis and there is no trivial way to collect time measurements from all of them. On the other hand, the runtime of functions often depends on their context of execution, which makes some of them unreliable for rootkit detection and sources of false alerts. With this paper we aim to overcome these challenges by answering the following research questions. \textit{RQ1: What system calls enable the observation of rootkits that hide files? RQ2: How can delays of relevant function calls be observed? RQ3: To what degree can anomaly detection techniques leverage system call function timings to uncover hidden rootkit activities?}

While investigating the topic of rootkit detection we noticed that one of the biggest problems that currently holds back research in the area of anomaly-based rootkit detection is the lack of data that can be used to evaluate rootkit detection approaches \cite{nadim2023kernel}. We therefore publish the data sets generated as part of our evaluation online\footnote{\url{https://zenodo.org/records/14679675}}. To facilitate reproduciblity of the results presented in this paper, we also publish the rootkit \footnote{\url{https://github.com/ait-aecid/caraxes}} as well as the implementation of our probing and detection mechanisms \footnote{\url{https://github.com/ait-aecid/rootkit-detection-ebpf-time-trace}} as open-source code. We hope that this will encourage others to extend our evaluation and generate even more public data sets for future research. We summarize our contributions as follows:

\begin{itemize}
	\item A framework and open-source implementation for kernel tracing with eBPF probes,
    \item an open-source rootkit and publicly available data sets of kernel function time measurements, and
    \item a detection mechanism for delayed function call timings.
\end{itemize}

The remainder of this paper is structured as follows. Section \ref{related} reviews rootkits and detection mechanisms. Section \ref{concept} outlines the concept of our approach. In Sect. \ref{tracing} we investigate relevant kernel functions and explain how we use probes for time measurement. In Sect. \ref{detection} we describe our approach for detection of anomalies. We evaluate our approach in Sect. \ref{eval} and discuss the results in Sect. \ref{discussion}. Finally, Sect. \ref{conclusion} concludes this paper. 

\section{Background \& Related Work} \label{related}

In this section we provide a review of existing open-source rootkits and discuss related works in the research area of rootkit detection.

\subsection{Rootkits} \label{rootkits}

This section explains the definition of a rootkit, enumerates common methods used by rootkits, and reviews open-source rootkits.

\subsubsection{Definition}

The term rootkit describes a software kit that provides root access to a system, which is the highest-privileged role on a system \cite{colbert2016cyber}. While rootkits by themselves are thus not inherently malicious \cite{bravo2011proactive}, they are often illicitly used as part of cyber attacks that allow adversaries to gain privileged access on a system without permission. After a rootkit has been deployed on a target system, it is often used to hide objects, such as files, processes, open ports, established connections, and the rootkit itself \cite{bravo2011proactive, singh2017detection}, to evade detection and enable attackers with continuous privileged access to prepare further attack steps such as information gathering \cite{tian2019kernel}. 

\subsubsection{Methods} \label{rkmethods}

The capability to effectively hide system objects and themselves from manual inspection and automatic detection systems is essential for rootkits to avoid that operators become aware of the intrusion, try to remove the rootkits, or disconnect the system from the network. There are several methods how rootkits interfere with systems and it is common to differentiate them based on the layer where the rootkit resides: kernel space, were all kernel activities are carried out with highest privileges, and user space, which is a lower-privileged domain containing user applications and libraries \cite{stuhn2024hidden}. One of the most straightforward methods used by user space rootkits is to exchange system binaries such as ``ls'' with modified versions that skip certain elements. This method has become outdated nowadays as it is trivial to detect by checksums \cite{bunten2004unix}. A modern alternative is to wrap functions in dynamically linked system libraries. In particular, many user space rootkits exploit the ``LD\_PRELOAD'' environment variable \cite{stuhn2024hidden}. 

There are several methods used by kernel space rootkits; we enumerate some of the most common ones in the following. (i) Loading the rootkit as a Loaded Kernel Modules (LKM). (ii) Leveraging extended Barkley Packet Filter (eBPF), which serves a similar purpose as LKM, but offers advantages such as higher stability and prebuilt hooks for system calls \cite{stuhn2024hidden}. (iii) Loading the rootkit into the initial file system of an operating system, which is loaded before the actual root file system is mounted \cite{landley2020ramfs}. (iv) Kernel patching, i.e., adding rootkit functionality to kernel source code, exchanging the kernel image, and forcing a reboot. (v) Containerization, which creates a name space, starts the init system inside a container, and leaves the rootkit outside of that container where it is not visible \cite{leibowitz2016horse}. (vi) Virtualization, which moves a running operating system into a hypervisor. As we show in the following, rootkits often use several of these methods in combination \cite{rutkowska2006introducing}.

\subsubsection{Open-source rootkits} \label{osrootkits}

\begin{table}[]
	\small
	\caption{Overview of the analyzed open-source rootkits}
	\label{tab:rootkits}
	\setlength{\tabcolsep}{4pt}
	\begin{tabular}{ccccc}
		\hline
		\makecell[c]{\textbf{Rootkit} \\ \textbf{Name}} & \makecell[c]{\textbf{Supported} \\ \textbf{Kernels}} & \makecell[c]{\textbf{Rootkit} \\ \textbf{Method}} & \makecell[c]{\textbf{Hooking} \\ \textbf{Mechanism}} & \makecell[c]{\textbf{Wrapped} \\ \textbf{Function}}  \\ \hline
		Puszek &  4.x & LKM & System call table & getdents  \\ \hline
		Suterusu &  2.6-3.x & LKM & \makecell[c]{Function pointer \\ in argument} & filldir  \\ \hline
		Diamorphine &  2.6-6.1 & LKM & kprobes & getdents  \\ \hline
		Reveng\_rtkit &  5.11 & LKM & kprobes & getdents  \\ \hline
		Reptile &  3.10-5.x & LKM & Binary patching & filldir  \\ \hline
		GLRK & 6+ & LKM & ftrace & -  \\ \hline
		Boopkit & 5.16+ & eBPF & eBPF & getdents \\ \hline
	\end{tabular}
\vspace{-7pt}
\end{table}

We review common open-source rootkits for the Linux operating system and analyze how they interact with the kernel. Table \ref{tab:rootkits} summarizes important features of the reviewed rootkits, including the function wrapped by the rootkit for the purpose of hiding. Note that most rootkits manipulate multiple system calls to achieve various goals and that our review solely focuses on hiding capabilities. \textbf{Puszek}\footnote{\url{https://github.com/Eterna1/puszek-rootkit}} is an LKM rootkit that first locates pointers to specific system calls in the system call table and then replaces the getdents system call to hide files. Instead of wrapping an entire system call, \textbf{Suterusu}\footnote{\url{https://github.com/mncoppola/suterusu}} replaces a function within the getdents system call, named filldir. Specifically, it does so when filldir is passed as a function pointer in one of the arguments of the context actor. \textbf{Diamorphine}\footnote{\url{https://github.com/m0nad/Diamorphine}} and \textbf{Reveng\_rtkit}\footnote{\url{https://github.com/reveng007/reveng\_rtkit}} are LKM rootkits that make use of \textit{kprobes}, a mechanism for debugging and tracing, to locate the system call table and wrap around the getdents system call. \textbf{Reptile}\footnote{\url{https://github.com/f0rb1dd3n/Reptile}} is another LKM rootkit that uses khook\footnote{\url{https://github.com/milabs/khook}} for binary patching of the filldir function; the mechanism behind this hacking tool is similar to kprobes. \textbf{Generic Linux Rootkit (GLRK)}\footnote{\url{https://codeberg.org/sw1tchbl4d3/generic-linux-rootkit}} makes use of \textit{ftrace}, a function tracer built into the Linux operating system, to execute code before and after function calls, which is equivalent to function hooking. However, the current implementation of the rootkit is designed for privilege escalation and does not support hiding; thus no function is wrapped. Finally, \textbf{Boopkit}\footnote{\url{https://github.com/krisnova/boopkit}} is an eBPF rootkit that enables process hiding, remote activation, and command execution. Specifically, the rootkit uses probes at system calls to skip the names of predefined process IDs when listed. Boopkit is another example of a rootkit that wraps the getdents system call for hiding.

\subsection{Rootkit Detection} \label{hids}

This section explains rootkit detection methods with a focus on learning-based detection approaches.

\subsubsection{Methods}

Given the diverse types of rootkits and the various ways how they interact with systems, it stands to reason that many different methods have been developed for rootkit detection. In their recent study, Nadim et al. \cite{nadim2023kernel} divide existing methods into six classes: (i) Signature-based methods make use of a predefined list of static signatures such as byte patterns that correspond to known rootkits \cite{bravo2011proactive}. Despite their simplicity, these methods pose a highly robust form of rootkit detection with low false positive rates, which is why most well-known host-based intrusion detection systems primarily rely on signatures. Unfortunately, they are unable to detect unknown rootkits for which no signatures exist \cite{bravo2011proactive} and can also be evaded by slightly modifying existing rootkits as well as mutating rootkits \cite{geetha2021nonvolatile}. (ii) Behavior-based detection aims to recognize actions of rootkits through anomalous states or behavior patterns observed in the operating system, such as unusual errors. (iii) Cross-view-based detection compares the visibility of the same objects on separate domains. Since rootkits often only hide objects on one domain, any divergences may indicate the presence of rootkits \cite{geetha2021nonvolatile}. For example, different outcomes obtained from enumerating kernel modules in the user space and searching loaded modules in memory may indicate a hidden rootkit. However, note that this method does not necessarily reveal the rootkit itself, but only some of its hidden objects \cite{bravo2011proactive}. (iv) Integrity-based detection recognizes changes to the kernel's static or dynamic data structures as opposed to recognizing the effects of such changes. In particular, changes of parts that are most often targeted by rootkits, such as patching of the system call table, are viable indicators for rootkit activity \cite{bravo2011proactive}. (v) Hardware-based detection leverages some of the other concepts mentioned in this enumeration but conducts analyses on an external device that cannot be accessed from the monitored host. (vi) Learning-based detection requires training data to capture a model that is subsequently used to classify unseen test data comprising data from both normal system behavior and rootkit activity. In the most common case, the training data only comprises normal instances and anomalies are detected as instances that deviate from the normal behavior model. Given that learning-based detection is the most relevant detection concept for this paper, we summarize existing works in this research area in the following.

\subsubsection{Learning-based Detection}

Learning-based detection requires data that is affected by rootkit activity and available in sufficient volumes to enable model training. Some authors have therefore turned to performance counters (HPC), which are special registers in microprocessors used for counting events \cite{sayadi2024intelligent}. Wang et al. \cite{wang2013numchecker} compare the event counts from normal and rootkit samples. Singh et al. \cite{singh2017detection} identify relevant HPCs through experimentation with five synthetic rootkits. Sayadi et al. \cite{sayadi2021towards} analyze HPCs as time-series. These approaches generally rely on machine learning methods such as support vectors machines (SVM), naive bayes classifiers, decision trees, and neural networks. Das et al. \cite{das2019sok} point out some downsides of HPCs, in particular, the need for expert knowledge to understand and collect HPCs correctly, diversity of HPCs across different processors, non-determinism of counters, and overcounting. Pattee et al. \cite{pattee2022performance} add that lack of documentation for HPCs, need for dedicated hardware, as well as energy constraints for machine learning in resource-limited devices pose issues for such detection approaches. Lu et al. \cite{lu2019data} point out that rootkits are able to mimic normal behavior patterns and recommend to focus on event timing, which is more difficult for them to replicate.

Several approaches therefore make use of event timing rather than count data for detection. Zimmer et al. \cite{zimmer2010time} detect buffer overflow attacks in cyber-physical systems by measuring timing bounds at specific checkpoints. Similarly, Salem et al. \cite{salem2016anomaly} use inter-arrival curves to estimate lower and upper bounds for event occurrence times. Lu et al. \cite{lu2019data} use lumped timing models to estimate limits of allowed time delays. Furthermore, they split up these timings into fine-granular subcomponent timing models to reduce influences of the operating system on time measurements and increase model accuracy. Carreon et al. \cite{carreon2020statistical} point out that lumped timing models suffer from high variability, which limits their ability to detect anomalies. The authors thus also focus on subcomponent timings and use the boundaries of cumulative distribution functions of time measurements to assign anomaly scores to unseen samples.

Some timing-based approaches specifically focus on system calls from standard operating systems. For example, Ezeme et al. \cite{ezeme2020framework} propose a framework that captures the order of system calls as well as their relative duration measured in CPU cycle counts and predict the expected counts for detection. Luckett et al. \cite{luckett2016neural} use neural networks to classify normal system and rootkit behavior based on system call timing. While the overall idea is similar to our work, they pursue supervised classification rather than detection and measure the runtime of entire system calls only. Another issue in their paper is that the method for data collection is not sufficiently described, which limits reproducibility of their results \cite{nadim2021review}. Dawson et al. \cite{dawson2018phase} also capture isolated system call timings with strace; specifically, they test their detection approach using the system calls \textit{open}, \textit{close}, \textit{read}, \textit{futex}, \textit{mmap2}, and \textit{clock\_gettime}. Brodbeck et al. \cite{brodbeck2012covert} measure system call latencies in mobile operating systems. Their findings suggest that rootkits can cause delays, but they do not analyze the complex distributions of system call timings in detail and also do not evaluate any detection algorithms. Contrary to these works, which capture the timing of entire system calls, our approach measures the inter-arrival timings of various functions within system calls, which allows us to isolate and reduce the effect of irrelevant factors such as the time needed to write and read from disk. Moreover, we are able to conduct our analyses on more fine-granular levels and capture even very slight time shifts, which can be of advantage when detecting rootkits that only modify these inner functions, such as Reptile (cf. Sect. \ref{osrootkits}). In addition, our detection method is designed for multimodal features that are prevalent in time measurements of system calls. We also investigate the influence of varying system conditions in detail. In the following, we outline the overall concept of our approach.

\section{Concept} \label{concept}

\begin{figure}
	\centering
	\includegraphics[width=\columnwidth]{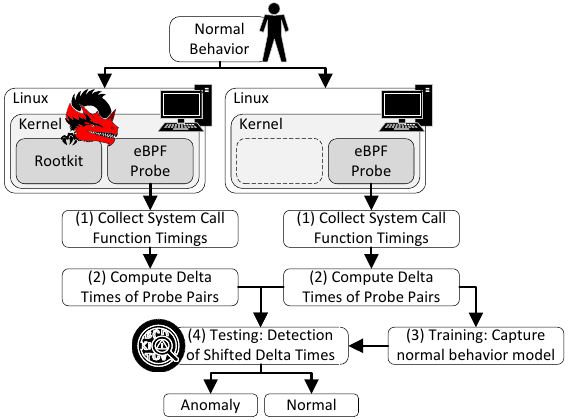}
	\caption{Overview of our concept that measures system call function timings to detect rootkits through delta time shifts.}
	\label{fig:concept}
	\vspace{-7pt}
\end{figure}

When kernel rootkits interfere with operating systems, they usually do so by wrapping around certain functions of system calls and thereby modifying the code executed at that position. As a consequence, the duration it takes to run the changed code is different in comparison to the runtime of the original code \cite{carreon2020statistical}. Specifically, the additionally executed code increases the overall runtime of the code block. The approach presented in this paper hinges on the assumption that function delta times, i.e., time intervals between certain positions in the executed code, increase sufficiently to enable differentiation between normal system behavior and rootkit activities.

Figure \ref{fig:concept} depicts an overview of our approach. Given that we aim to recognize rootkit activity through changed function delta times, we need to collect a baseline of measurements from an operating system that is free from rootkits and compare them to measurements from a system affected by rootkits. In both systems, we assume that normal user interactions take place continuously and that some of these executed commands trigger system calls affected by the rootkit, e.g., commands that enumerate files. As part of our approach, we inject probes into the kernel to capture execution times of inner functions of system calls. In particular, for a predefined set of functions that are likely to be affected by rootkit behavior, the probes collect and store the absolute time stamps of entering and returning from those functions as depicted in step (1) in Fig. \ref{fig:concept}. We outline kernel tracing and probe injection in Sect. \ref{tracing}. Note that measurements are stored separately for the normal and rootkit case; in practice it is obviously not straightforward to make this differentiation and collect clean data sets, however, we argue that the detection of changed system call function timings in comparison to any past system state can already be useful indication for potential rootkit activity. We discuss practical and online applicability of our approach in Sect. \ref{online} and Sect. \ref{discussion} in more detail.

Step (2) computes the delta times between pairs of time stamps collected in step (1). To this end we consider two strategies: (i) computation of delta times between entry and return point of certain functions, and (ii) computation of delta times between two subsequently encountered probes independent from the respective function. We discuss each method in Sect. \ref{measurement}. In step (3), we then apply machine learning to classify unseen delta times and detection of rootkits. Note that we pursue semi-supervised detection, meaning that a portion of only the normal delta times is used to generate a model of normal behavior; all unseen samples that deviate too much from that model are detected as anomalies that are potentially caused by rootkit activities. Our method is suitable for offline detection, where data is forensically analyzed to differentiate between normal and rootkit samples, and online detection, where the model is incrementally updated and recognizes unusual changes of delta times on-the-fly. The following sections explain our methods for data collection and outlier detection in detail.

\section{Kernel Tracing} \label{tracing}

In this section we discuss the relevance of functions in system calls and describe our method to inject probes for time measurement.

\subsection{Analysis of Relevant System Call Functions} \label{systemcallfunc}

Log data is commonly used for anomaly detection in the cyber security domain \cite{khraisat2019survey}. Unfortunately, rootkits have the highest privileges on a system and are thus able to manipulate the contents and generation of log data in such a way that their presence remains hidden, e.g., by suppressing certain log messages. Even though rootkits may alter the entire system at will, they cannot easily replicate the system behavior as if they were not present on the system, since any action that they perform still needs to be executed by the kernel, which is where they leave detectable traces. 

One way to monitor kernel activities is to analyze system calls (syscalls), which are an interface for user programs to request resources and interact with the kernel. Most operating systems offer hundreds of system calls\footnote{\url{https://man7.org/linux/man-pages/man2/syscalls.2.html}}, with some of the most common ones being open, read, write, and fork. Some modern host-based intrusion detection systems are capable of monitoring single invocations of system calls and sequences of system calls have long been used for malware detection \cite{forrest2008evolution}; however, rootkits do not necessarily affect multiple system calls, but may only affect a single system call or just one of its inner functions. As a consequence, existing intrusion detection systems do not monitor kernel activity in sufficient granularity to uncover rootkit activities. 

\begin{figure}
  \begin{lstlisting}[style=mypseudo,escapechar=§]
           | x64_sys_call() {
           |   __x64_sys_getdents64() {
           |     __fdget_pos() {
  0.476 us |       __fget_light();
           |       mutex_lock();
  3.255 us |     }
           |     iterate_dir() { §\label{line:iteratedir}§
           |       security_file_permission();
           |       down_read_killable();
           |       dcache_readdir() { §\label{line:dcachereaddir}§
           |         filldir64() { §\label{line:filldir}§
  0.699 us |           verify_dirent_name(); §\label{line:verifydirent}§
  1.727 us |         } §\label{line:filldirret}§
  0.458 us |         _raw_spin_lock();
  0.466 us |         _raw_spin_unlock();
           |         filldir64() { §\label{line:filldir2}§
  0.544 us |           verify_dirent_name();§\label{line:verifydirent2}§
  1.469 us |         } §\label{line:filldirret2}§
           |         scan_positives() {
  0.462 us |           _raw_spin_lock();
  0.461 us |           _raw_spin_unlock();
  0.469 us |           dput();
  4.787 us |         }
                     /** loop **/
  0.473 us |         dput();
118.077 us |       }
           |       touch_atime() { §\label{line:touchatime}§
           |         atime_needs_update() {
  0.464 us |           make_vfsuid();
  0.466 us |           make_vfsgid();
           |           current_time();
  4.233 us |         }
  5.222 us |       }
  0.498 us |       up_read();
132.216 us |     }
           |     __f_unlock_pos();
138.769 us |   }
139.968 us | }
  \end{lstlisting}
  \caption{Excerpt from the getdents call stack and function timings collected with the function tracer (ftrace).}
  \label{code:call_stack}
  \vspace{-7pt}
\end{figure}

Due to the fact that many system calls exist and each of them involves a myriad of functions, making a reasonable selection for monitoring is vital to limit the amount of data to be analyzed and focus on those points that are most likely targeted by rootkits. Rather than manually sifting through all available system calls to make this decision, we reviewed seven open-source rootkits (cf. Sect. \ref{rootkits}) and found that every single one of them makes use of the \textit{getdents} system call or one of its inner functions to enable hiding of files. We also noticed that getdents is explicitly mentioned in technical reports on cyber attacks involving rootkits \cite{remillano2019skidmap}. This is intuitively reasonable since getdents is the only interface for a user program to list the contents of a directory in Linux, which is an interface that rootkits needs to control in order to hide objects, such as files or themselves \cite{brodbeck2012covert}. We display a shortened version of the getdents system call in Fig. \ref{code:call_stack}, which we generated with the function tracer\footnote{\url{https://www.kernel.org/doc/html/latest/trace/ftrace.html}} (ftrace). Note that we omit irrelevant functions from the code for brevity and that executed functions may differ depending on the context in which getdents is invoked. As visible in the figure, the function tracer also provides time measurements that describe how long it took to complete single functions. These timings are the primary data source for our detection algorithm. In order to collect time measurements in a structured way, we inject probes into the kernel, which we describe in the following section.

\subsection{Injection of eBPF Probes} \label{injection}

\begin{figure}
  \begin{lstlisting}[style=myC,escapechar=§]
BPF_RINGBUF_OUTPUT(buffer, 1 << 4);§\label{line:ringbuf}§
struct event {
    unsigned long time;
    u32 pid;
    u32 tgid;
};
int probe(struct pt_regs *ctx) {
    struct event *event = buffer.ringbuf_reserve(sizeof(struct event));
    if (!event) {return 1;}
    u64 pid_tgid = bpf_get_current_pid_tgid(); §\label{line:get_pid_tgid}§
    event->tgid = pid_tgid >> 32; §\label{line:get_tgid}§
    event->pid = (u32) pid_tgid; §\label{line:get_pid}§
    event->time = bpf_ktime_get_ns(); §\label{line:get_time}§
    buffer.ringbuf_submit(event, 0); §\label{line:rbuf_submit}§
    return 0;
}
  \end{lstlisting}
  \vspace{-14pt}
  \caption{Implementation of our eBPF probe.}
  \label{code:eBPF-probe}
  \vspace{-7pt}
\end{figure}

While the function tracer is a viable method to obtain function time measurements, we opt for the more modern extended Berkeley Packet Filter (eBPF) to implement time measuring probes. The eBPF is a Linux kernel technology that enables developers to build programs that run securely in kernel space. Figure \ref{code:eBPF-probe} shows our implementation of an eBPF probe. Each probe first obtains identifiers for the current process (pid) and thread group (tgid) in Line \ref{line:get_pid_tgid}, which are stored in the respective variables in Lines \ref{line:get_pid} and \ref{line:get_tgid}. Then, the current time is measured in nanoseconds in Line \ref{line:get_time}. Finally, these variables are written as an event in the ring buffer (Line \ref{line:rbuf_submit}), which is created in Line \ref{line:ringbuf} and managed by the BPF Compiler Collection\footnote{\url{https://github.com/iovisor/bcc}} (BCC) that is also used to inject the probes. 

BCC supports injection at enter or return points of all functions that can be traced by eBPF. For each probe, we therefore obtain two measurements that we refer to as probe-enter and probe-return respectively, where we use the name of the function to refer to the probe. For example, for function filldir in Fig. \ref{code:call_stack}, we obtain time measurements from probes filldir64-enter (Lines \ref{line:filldir} and \ref{line:filldir2}) and filldir64-return (Lines \ref{line:filldirret} and \ref{line:filldirret2}). To avoid that our script collecting the events from BCC influences time measurement by triggering system calls when storing the data, all events are held in memory until all probes are unloaded. Note that some functions cannot be traced by eBPF and are thus neglected for our analyses.

\section{Detection} \label{detection}

This section outlines our detection algorithm. We explain two strategies to derive delta times from time measurements and then describe an approach to detect anomalies based on shifted delta times.

\subsection{Computation of Delta Times} \label{measurement}

The main idea behind our detection approach is that it takes more time to execute code of functions that are wrapped by the rootkit in comparison to executing the original functions without any additions made by the rootkit \cite{carreon2020statistical}. To capture the duration of time interval between any two probes $p1$ and $p2$ based on the absolute time measurements collected as described in the previous section, it is necessary to subtract the time stamp of an event observed at probe $p_2$ with the time stamp of the event observed at probe $p1$ that chronologically occurs before. In the following, we use the colon to denote this pairing of probes, i.e., $p_1$:$p_2$. For example, filldir64-enter:verify\_dirent\_name-enter are delta times between the probes at filldir64-enter (Lines \ref{line:filldir} and \ref{line:filldir2} in Fig. \ref{code:call_stack}) and verify\_dirent\_name-enter (Lines \ref{line:verifydirent} and \ref{line:verifydirent2}). Note that processes are often running in parallel and their workflows interleave, causing that chronological sorting alone is not sufficient to correctly pair probes. We therefore first split all measurements into groups by process identifier (pid) and then sort all time stamps in each group in ascending order before subtraction.

\begin{figure}
	\centering
	\includegraphics[width=\columnwidth]{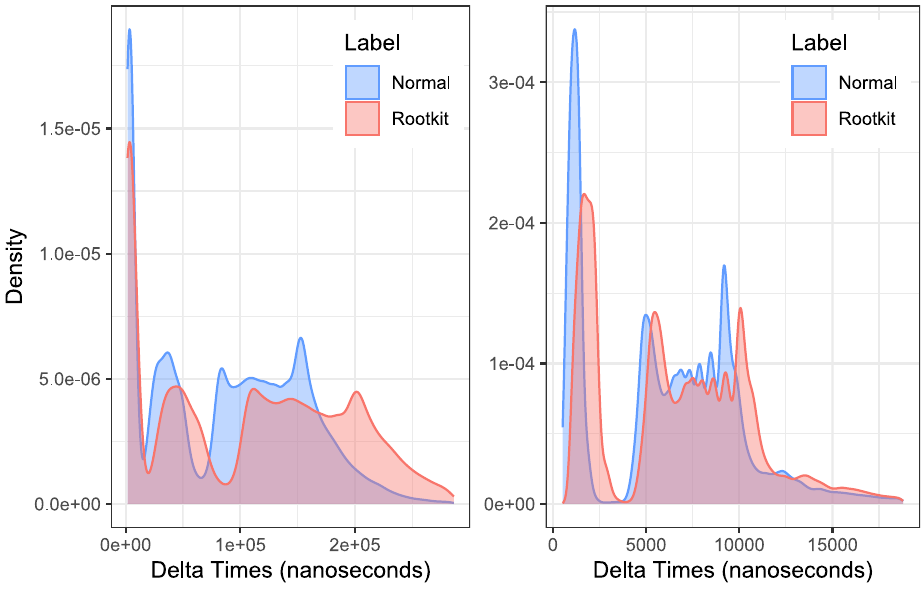}
	\caption{Time measurements at probes iterate\_dir-enter:iterate\_dir-return (left) and filldir64-return:filldir64-enter (right) depicting longer delta times when the rootkit is active (red) in comparison to normal system behavior (blue).}
	\label{fig:sample}
	\vspace{-7pt}
\end{figure}

To reduce the immense number of possible combinations of probes to a manageable amount, we propose two strategies for probe pair selection. The first strategy only considers enter and return probes injected at the same function and subtracts the most recent time stamp of a probe-enter event from the probe-return event, effectively measuring the time it takes to execute that function. Accordingly, we refer to this strategy as function-grouping in the following. Note that the described procedure is repeated for each event from every probe. The second strategy sequentially iterates through all events in chronological order and subtracts the time stamp from the latter event with the time stamp from the former event, independent from the probe and its corresponding function. We refer to this strategy as sequence-grouping. Each strategy has their own benefits. While sequence-grouping captures short delta times between adjacent functions, which enables fine-granular analysis, function-grouping captures entire functions and may thus be better suited to detect rootkits that wrap around these functions. 

Figure \ref{fig:sample} depicts two illustrative distributions of delta times, where the left one corresponds to probes iterate\_dir-enter:iterate\_dir-return using function-grouping and the right one corresponds to filldir64-return:filldir64-enter using sequence-grouping. We select these samples, because they both show a visible delay of delta times when the rootkit is active in comparison to delta times collected during normal system operation. In the following section, we describe a mechanism that automatically detects these shifts.

\subsection{Shift Detection} \label{shift}

The main idea of our detection approach is to automatically recognize shifts in delta times that are computed as described in the previous section. However, the density curves plotted in Fig. \ref{fig:sample} reveal that even though there is a visually apparent shift in the distributions of delta times, it is non-trivial to measure that shift and determine what degree of shift is still tolerable and likely a product of natural variation rather than an actual change in the code caused by the rootkit. In particular, delta times from some probes may be influenced by system conditions and thus too volatile for use as a baseline. To be able to assess the amount of expected variation, we therefore assume to have several batches of data at our disposal, where each batch was taken independently over a certain period of time and contains sufficiently many delta times to estimate their distribution at that point in time. This allows us to apply statistical tests on the data in spite of varying system conditions and noise. In the following, we refer to batches collected during normal system operation and batches collected while a rootkit is active on the system as normal and rootkit batches respectively.

Due to the fact that the distributions of delta times are multimodal (i.e., involve multiple local peaks) and contain outliers, it is not feasible to simply rely on the mean and standard deviation for statistical testing \cite{luckett2016neural}. To overcome this issue, we compare the quantiles of delta time distributions from different batches, because quantiles accumulate close to peaks where robust measurements of delta time shifts are possible. For example, a natural way of determining the shift for the distributions in Fig. \ref{fig:sample} is to measure the horizontal offset at some prominent peaks. Moreover, quantiles enable to determine whether only a portion of the delta times has been shifted, which can be relevant if only some invocations of the same function are affected by the rootkit.

We use equidistant spacing between $0$ and $1$ for a predefined number of quantiles $q$ to capture most of the entire distribution while at the same time ensuring robustness against outliers. For example, for the trivial case of $q=1$, the median (or $0.5$-quantile) will be used, while for $q=4$, the $0.2$-, $0.4$-, $0.6$-, and $0.8$-quantiles will be used. We combine the quantiles of delta times for a set of normal batches in a training set $V_{p_1:p_2}$ of size $n \times q$, where $n$ is the number of training batches and $q$ is the number of quantiles. For an unseen batch of delta times $x$, which can be from the remaining normal batches or one of the rootkit batches, we compute the squared Mahalonobis distance using Eq. \ref{mhd}, where $\mu(V_{p_1:p_2})$ is the mean of delta times for each quantile in $V_{p_1:p_2}$ and $\Sigma^{-1}$ is the inverse covariance matrix of $V_{p_1:p_2}$. We then compute the $p$-value for a specific $p_1$:$p_2$ using a $\chi^2$-test with $q$ degrees of freedom as stated in Eq. \ref{chisq}, where $cdf$ is the cumulative distribution function \cite{etherington2019mahalanobis}. 
\begin{equation} \label{mhd}
	D_{p_1:p_2}^2 =\left( x_{p_1:p_2} - \mu(V_{p_1:p_2}) \right)^\top \Sigma^{-1} \left( x_{p_1:p_2} - \mu(V_{p_1:p_2}) \right)
\end{equation}
\begin{equation} \label{chisq}
	p\text{-value}_{p_1:p_2} = 1 - \chi^2(cdf(D_{p_1:p_2}^2, q))
\end{equation}
Low $p$-values close to $0$ indicate that delta times at one or multiple quantiles are significantly shifted from the expected means considering the variations observed in the training data, while $p$-values close to $1$ indicate the opposite. Note that this method is semi-supervised, because we assume that our training set only contains normal data and is free from anomalies introduced by rootkits, while test data may contain batches from both classes. 

The aforementioned computation of $p$-values is applicable to a single combination of probes $p_1$:$p_2$, independent from the grouping strategy used. However, many functions within the getdents system call could be targeted by rootkits, and it is not possible to know beforehand on which combination of probes to focus on. We therefore propose to involve as many probes as possible for shift detection and combine their respective $p$-values. Given that rootkits can wrap any function, i.e., only affect delta times collected from a single probe, we consider it sufficient if one of the $p$-values is below a certain threshold $\theta$ to detect the entire batch as an outlier that potentially indicates rootkit activity. The sample is only considered normal if all $p$-values are above the threshold. In the following section, we evaluate the effectiveness of this detection method by deploying a rootkit on a real kernel and analyzing the delta times.

\section{Evaluation} \label{eval}

This section covers the evaluation of our work. We introduce a novel open-source rootkit and describe the generated public data sets, which we use to evaluate our detection approach.

\subsection{CARAXES: A Cyber Analytics Rootkit}

We initially planned to select one of the open-source rootkits reviewed as part of Sect. \ref{rootkits} for our evaluation; however, we realized that none of these rootkits is able to run on modern kernels. For example, Diamorphine overwrites the system call table to perform system call wrapping, which is not possible since Linux kernel version 6.9 started removing the system call table as a security measure to avoid speculative execution. Reptile wraps around the filldir function rather than an entire system call, but this function has been changed in Linux kernel version 6.1. In addition, Reptile relies on the function ``kallsyms\_lookup\_name'', which is not available in Linux kernels above version 6. Similar issues exist for other rootkits. An exception to this observation is GLRK, which works without issues but unfortunately does not support file hiding.

\begin{figure}
	\begin{lstlisting}[style=myC,escapechar=§]
static bool hook_filldir64(struct dir_context     *ctx, const char *name, int namlen, loff_t offset, u64 ino, unsigned int d_type) {
	struct readdir_callback *buf = container_of(ctx, struct readdir_callback, ctx);
	if (strstr(name, MAGIC_WORD)) { §\label{line:magic}§
		buf->result = -ENOENT;
		return false;
	}
	return orig_filldir64(ctx, name, namlen, offset, ino, d_type);§\label{line:orig}§
}
	\end{lstlisting}
	\vspace{-14pt}
	\caption{Wrapper for the filldir function.}
	\label{code:rootkit}
	\vspace{-7pt}
\end{figure}

To overcome this problem, we present a \textit{Cyber Analytics Rootkit for Automated and eXploratory Evaluation Scenarios} (CARAXES). We base our implementation on GLRK, which we extend with functionality to hide files if they contain specific keywords in their names or belong to a certain user or group. Process hiding is implicitly supported by specifying user and group identifiers of processes to be hidden. We implement two different ways of hooking into the kernel: wrapping the getdents system call (entire code of Fig. \ref{code:call_stack}) and wrapping the filldir function (Lines \ref{line:filldir} and \ref{line:filldir2} in Fig. \ref{code:call_stack}) within that system call. We noticed during our experiments that our probing mechanism (cf. Sect. \ref{injection}) prevents the rootkit from wrapping getdents; for this reason, we focus on filldir wrapping for our evaluation. Figure \ref{code:rootkit} displays the filldir wrapper, which checks in Line \ref{line:magic} if the name of an enumerated file contains the keyword \textit{MAGIC\_WORD} and skips the file in case of a match, rendering it invisible. Otherwise, the wrapper passes all parameters to the original filldir function in Line \ref{line:orig} and returns its result.

CARAXES is designed for modern Linux kernel versions and tested on version 6.11. We also make CARAXES publicly available as open-source to allow others to generate new data sets and extend the rootkit with additional features (cf. Sect. \ref{intro}). 

\subsection{Data Generation} \label{setup}

This section outlines our procedure for data generation and explains which scenarios we consider for system variations.

\subsubsection{Procedure} \label{procedure}

\begin{figure}
	\centering
	\includegraphics[width=\columnwidth]{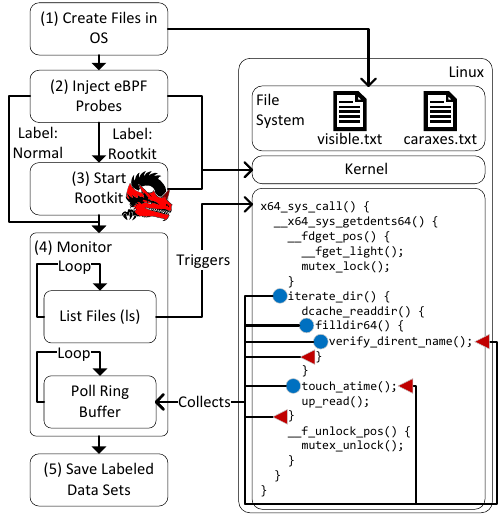}
	\caption{Overview of our data generation procedure.}
	\label{fig:eval}
	\vspace{-7pt}
\end{figure}

To generate the data sets that we use to evaluate our probing framework and detection approach, we set up a data generation procedure involving the rootkit described in the previous section. Figure \ref{fig:eval} visually summarizes our setup, which consists of the following steps: (1) We create a directory that contains two files; one arbitrary file and one that should be hidden by the rootkit. We identify the file to be hidden through its name, i.e., the rootkit is programmed to hide all files containing the keyword ``caraxes'' in its name. (2) We inject the eBPF probes using our framework from Sect. \ref{tracing}. For the purpose of this evaluation, we select the functions iterate\_dir (Line \ref{line:iteratedir} in Fig. \ref{code:call_stack}), filldir (Lines \ref{line:filldir} and \ref{line:filldir2}), verify\_dirent\_name (Line \ref{line:verifydirent}), and touch\_atime (Line \ref{line:touchatime}) due to their proximity to the affected filldir function. (3) We then either activate the rootkit or proceed without an active rootkit; this decision determines the label of the resulting data set. (4) We then start a loop that repeatedly executes the ``ls'' command to list files in the previously generated directory containing the two files, which performs getdents system calls that involve the wrapped function. The command is executed $100$ times without any waiting time in between. In parallel, we run another loop that continuously polls the ring buffer to collect time measurements from probes. (5) After completion, the measurement times are written from memory to the file system, including meta-information such as labels and parameters of the generation procedure.

We refer to the data set resulting from this procedure as one batch of data. We run this procedure many times to collect a sufficiently large amount of normal and rootkit batches to train and test our approach. Between each batch, we pause the iteration for 10 seconds using the sleep command in order to ensure that there are no artifacts from previous batches introduced during data collection. As a side effect, this causes that the collected data spans over a longer duration, which allows to analyze whether the data is influenced by trends and concept drift over time.

\subsubsection{Scenarios} \label{scenarios}

In addition to noise, trends, and concept drift originating from the system, parameters of the evaluation setup need to be taken into consideration when analyzing the data. To investigate the influence of these factors, we carry out the generation of normal and rootkit batches in five different scenarios: (i) \textit{Default}. The procedure is executed as described in Sect. \ref{procedure}. (ii) \textit{File Count}. Other than in the default scenario where one normal file and another file to be hidden are generated, this scenario involves a random selection of 10 to 100 files of each type. (iii) \textit{Filename Length}. In contrast to the default scenario where names of files consist of at most 8 random characters (and the keyword ``caraxes'' for files to be hidden), the lengths of file names in this scenario are randomly selected in the range of 20 to 60 characters. (iv) \textit{ls-basic}. We replace the ``ls'' command that is used to enumerate files with a custom implementation named ``ls-basic''\footnote{\url{https://github.com/ait-aecid/rootkit-detection-ebpf-time-trace/blob/main/ls-basic.c}} that does not rely on any libraries and allows us to know exactly which system calls are invoked. (v) \textit{System Load}. We run the tool stress-ng\footnote{\url{https://github.com/ColinIanKing/stress-ng}} in background while collecting the data to simulate a system under load. The data sets are generated on Ubuntu 22.10 with Linux Kernel 5.19, 32GB RAM, and 8 vCPUs.

\subsection{Data Sets}

This section provides an overview of three data sets published alongside this paper: one data set of time measurements at probes and two data sets of delta times derived from these time measurements.

\subsubsection{Time Measurements at Probes} \label{data}

\begin{table}
\small
\caption{Overview of the generated data sets}
\label{tab:datasets}
\begin{tabular}{cccccc}
\hline
\textbf{Scenario} & \textbf{Label} & \textbf{\makecell[c]{Start \\ time}} & \textbf{\makecell[c]{End \\ time}} & \textbf{Batches} & \textbf{\makecell[c]{\#Events \\ (median)}} \\ \hline
\multirow{2}{*}{Default} & Normal & 10:58:25 & 11:44:24 & 150 & 39538 \\ \cline{2-6} 
 & Rootkit & 11:44:44 & 12:17:18 & 100 & 52524 \\ \hline
\multirow{2}{*}{File Count} & Normal & 12:17:37 & 13:05:46 & 150 & 78016 \\ \cline{2-6} 
 & Rootkit & 13:06:06 & 13:40:04 & 100 & 89368 \\ \hline
System & Normal & 13:40:24 & 14:29:06 & 150 & 39274 \\ \cline{2-6} 
Load & Rootkit & 14:29:27 & 15:03:56 & 100 & 52115 \\ \hline
\multirow{2}{*}{ls-basic} & Normal & 15:04:14 & 15:49:45 & 150 & 33230 \\ \cline{2-6} 
 & Rootkit & 15:50:04 & 16:22:15 & 100 & 39818 \\ \hline
Filename & Normal & 16:22:33 & 17:08:35 & 150 & 39544 \\ \cline{2-6} 
Length & Rootkit & 17:08:55 & 17:41:30 & 100 & 52120 \\ \hline
\end{tabular}
\vspace{-7pt}
\end{table}

Table \ref{tab:datasets} summarizes the first data set of probe measurements collected as described in Sect. \ref{injection}. The table differentiates between normal and rootkit batches as well as the scenario in which they are collected (cf. Sect. \ref{scenarios}). In addition, we also provide the start and end times of collection in the table, which shows that we collected the data successively by iterating through each scenario and alternating between normal and rootkit cases. For each scenario, we collect $150$ normal and $100$ rootkit batches. The reason for this is that we intend to use $50$ batches (a third of the normal data) for training and thereby leave a balanced test set of $100$ normal and $100$ rootkit batches respectively. The last column of the table states the median number of events per batch, which shows that event counts vary across scenarios and that rootkit batches involve more events compared to normal batches of the same scenario. 

\begin{figure}
\centering
\includegraphics[width=\columnwidth]{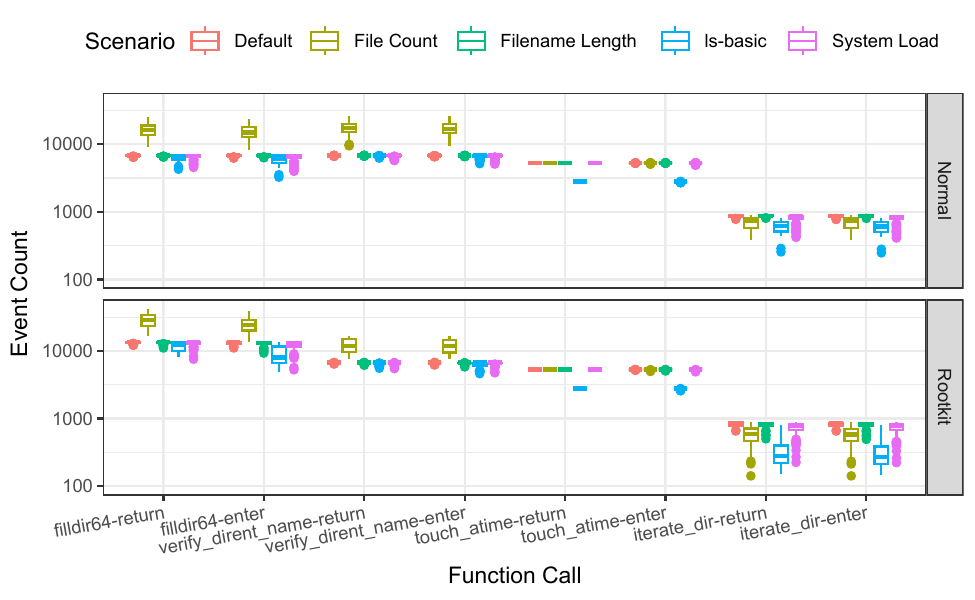}
\caption{Number of events per probe across scenarios.}
\label{fig:events}
\vspace{-7pt}
\end{figure}

Figure \ref{fig:events} provides a more detailed view in the number of events per batch by additionally separating the counts by probe. The figure reveals several interesting aspects. First, the number of time measurements varies across probes. For example, the iterate\_dir function is invoked less frequently than other probed functions, causing that fewer time stamps are collected from the probes of that function. This is simply explained by the program workflow that calls some functions more often than others, e.g., filldir is invoked multiple times within the iterate\_dir function (cf. Fig. \ref{code:call_stack}). Second, the enter and return probes of each function yield roughly the same number of measurements. While there are some outliers indicating that few measurements have not been successfully retrieved from the ring buffer, the overall distributions suggest that most events have been collected and all probes have been adequately captured. Third, event counts collected at probes of different functions vary across scenarios. For example, increasing the number of files triggers more invocations of the filldir and verify\_dirent\_name functions. Fourth, there are some differences in the number of events depending on whether a rootkit is active or not, which explains the event counts stated in Table \ref{tab:datasets}. The most significant divergence occurs with the filldir function, which is reasonable since the rootkit injects a new function at the same position that also invokes the original function, which increases the total number of collected measurements at that probe. 

The latter observation implies that in our data set the number of event counts can be used as a trivial way of detecting rootkits. However, this is not necessarily the case in general since rootkits may be injected in the kernel in many different ways (cf. Sect. \ref{rkmethods}); some of which do not trigger an additional function call. For example, rootkits may substitute the targeted function entirely by replacing the kernel with a modified version and forcing a reboot. Analysis of function timing is also capable of detecting changes within the code and is not limited to function calls. In addition, while detection based on event counts fails if the specific function targeted by the rootkit is not probed, timing-based detection is still able to recognize delays as long as the affected function is invoked within the probed function, causing the entire runtime of the outer function to increase. Finally, function timing is non-trivial to replicate for rootkits that try to mimic normal behavior in comparison to other features collected from the kernel \cite{lu2019data}. In the following, we therefore only focus on function timing for detection.

\subsubsection{Delta Times}

\begin{figure}
	\centering
	\includegraphics[width=\columnwidth]{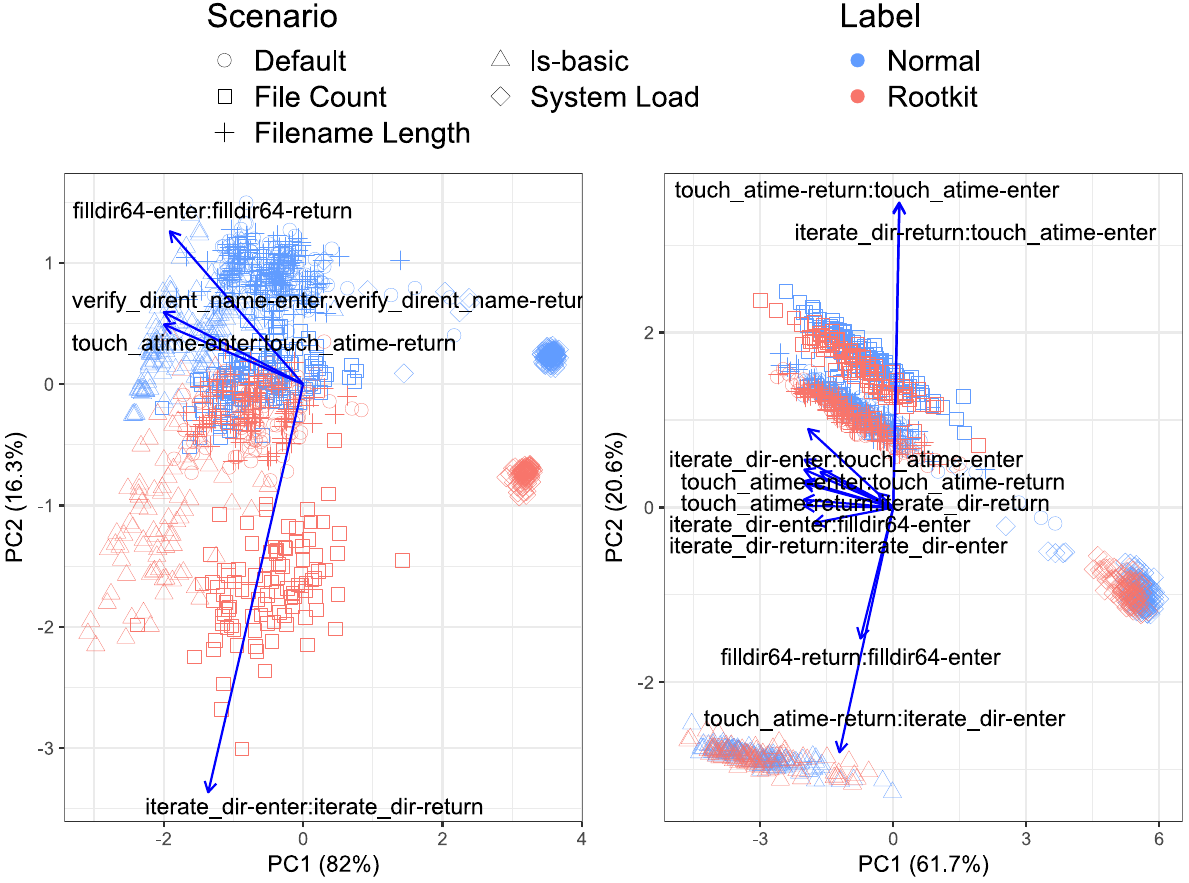}
	\caption{Biplots of delta times computed via function-grouping (left) and sequence-grouping (right).}
	\label{fig:pca}
	\vspace{-7pt}
\end{figure}

Based on the data set of time measurements described in the previous section, we compute two data sets of delta times using function-grouping and sequence-grouping following the strategies outlined in Sect. \ref{measurement}. To visualize the delta times, we compute the median value of delta times for each pair of probes in every batch and apply principal component analysis (PCA). Figure \ref{fig:pca} shows biplots of the first two principal components for delta times computed using function-grouping (left) and sequence-grouping (right), where symbols indicate the collection scenario (cf. Sect. \ref{scenarios}) and color differentiates normal (blue) from rootkit (red) batches. It is apparent in the left plot that many batches corresponding to rootkit behavior exhibit increased delta times at probes iterate\_dir-enter:iterate\_dir-return, which is also the combination of probes we depict for the default scenario in the left side of Fig. \ref{fig:sample}. While this suggests that many these batches can be correctly detected as outliers, it seems difficult to discern batches corresponding to different scenarios, with the exception of the system load scenario where batches are far away from the others and both classes are well separable. The plot on the right hand side suggests that sequence-grouping performs better at separating batches from different scenarios, with the exception of batches corresponding to the scenario where filename lengths are varied that mostly overlap with batches from the default scenario. Contrary to the plot on the left hand side, however, rootkit batches are not as simple to separate from normal batches. In the following, we evaluate our detection approach with both strategies and compare the results.

\subsection{Results}

This section presents the results of our evaluation. We first highlight that delta times are shifted at certain probes, which indicates rootkit activity. We then apply our detection algorithm and evaluate its ability to detect rootkit from normal activity given a training set of only normal data.

\subsubsection{Delta Time Shifts}

While the biplots presented in the previous section provide a rough indication about the probes that are suitable to detect rootkits based on function timing, they do not communicate precise delays and combine the influence of several probes. Moreover, while we only use the median for PCA, we aim to involve multiple quantiles for detection (cf. Sect. \ref{shift}). Inspired by the shift function that has been used to compare which parts of distributions are shifted \cite{doksum1974empirical, wilcox2012comparing}, we subtract each quantile of delta times collected from rootkit batches with those of normal batches. In the following, we always use nine quantiles ($q=9$) for our analyses and detection evaluations.

\begin{figure}
	\centering
	\includegraphics[width=\columnwidth]{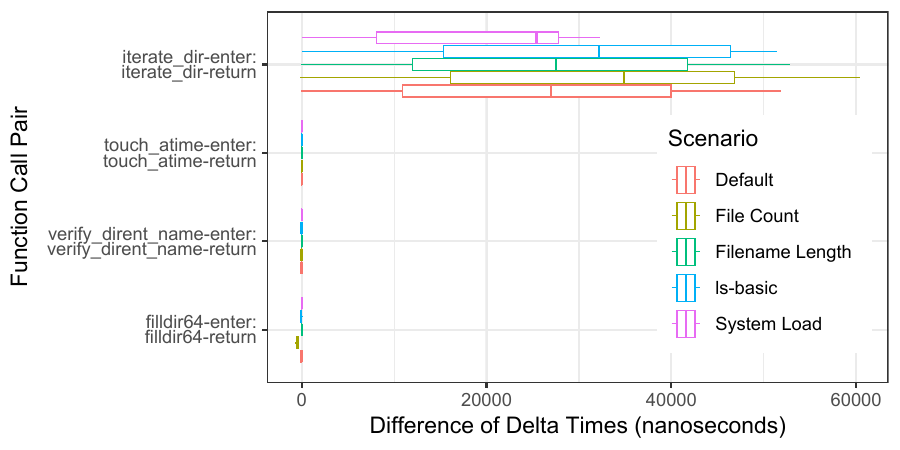}
	\caption{Delta time shifts for function-grouping.}
	\label{fig:shifts_name_fun}
	\vspace{-7pt}
\end{figure} 

\begin{figure}
\centering
\includegraphics[width=\columnwidth]{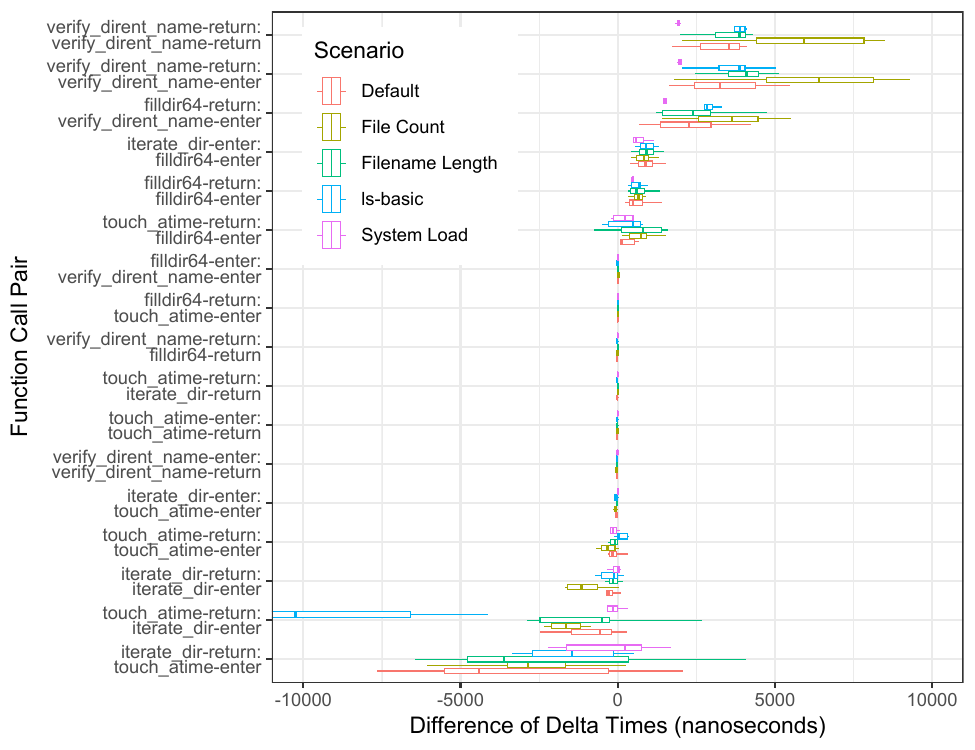}
\caption{Delta time shifts for sequence-grouping.}
\label{fig:shifts_name_seq}
\vspace{-7pt}
\end{figure}

Figure \ref{fig:shifts_name_fun} depicts the differences of delta times for function-grouping as a boxplot, separated by probes and scenarios. As visible in the plot, only the iterate\_dir-enter:iterate\_dir-return probes show a significant shift across all scenarios; other probes seem mostly unaffected by the influence of the rootkit. While the shifts spread over a wide range, these results suggest that most of the batches collected during rootkit activity can be detected. In contrast, Fig. \ref{fig:shifts_name_seq}, which depicts the delta times computed with sequence-grouping, show significantly more diverse patterns. While there is hardly any shift noticeable at some probes in the middle of the plot and some probes in the bottom of the plot show no obvious trends, the probes depicted in the top of the plot clearly indicate shifted delta times. While some of them exhibit a rather high variation for certain scenarios, we find that iterate\_dir-enter:filldir64-enter and filldir64-return:filldir64-enter are good indicators as the shifts seem relatively constant and have little variation across quantiles, batches, and scenarios. For this reason, we select filldir64-return:filldir64-enter as an illustrative example to display time shifts in the right side of Fig. \ref{fig:sample}. In the following, however, we do not make any manual selections, but instead use the entirety of the data to leverage as much information as possible and facilitate realistic evaluation.

\subsubsection{Offline Detection} \label{offline}

To evaluate our detection algorithm described in Sect. \ref{shift}, we split the batches of each of our two data sets of delta times into training and test data sets. As proposed in Sect. \ref{data}, we use delta times of $50$ normal batches for training and leave $100$ remaining normal and $100$ rootkit batches for testing. We thereby pursue evaluation in an offline setting, i.e., sample the training data randomly from the normal batches and repeat the sampling $100$ times so that we are able to estimate the variance of our results. Note that we evaluate the detection performance separately for every scenario, i.e., we sample the training data only for normal data from one specific scenario and then evaluate detection using the test data of that scenario, and repeat that for every scenario independently. The reason for this is that we do not assume that our normal behavior model generated from data of one scenario is capable of differentiating normal and rootkit batches from another scenario. We thus count true positives ($TP$) as rootkit batches that are detected as anomalous by our approach, false positives ($FP$) as normal batches that are detected as anomalous, false negatives ($FN$) as rootkit batches that are not detected as anomalous, and true negatives ($TN$) as normal batches that are not detected as anomalous. We sum up all of these counts for every scenario and compute the true positive rate or recall ($TPR = Rec = \frac{TP}{TP+FN}$), true negative rate ($TNR = \frac{TN}{TN+FP}$), precision ($Prec = \frac{TP}{TP+FP}$), accuracy ($Acc = \frac{TP + TN}{TP + TN + FP + FN}$), and F1 score ($F1 = \frac{2 \cdot Prec \cdot Rec}{Prec + Rec}$).

\begin{figure}
	\centering
	\includegraphics[width=\columnwidth]{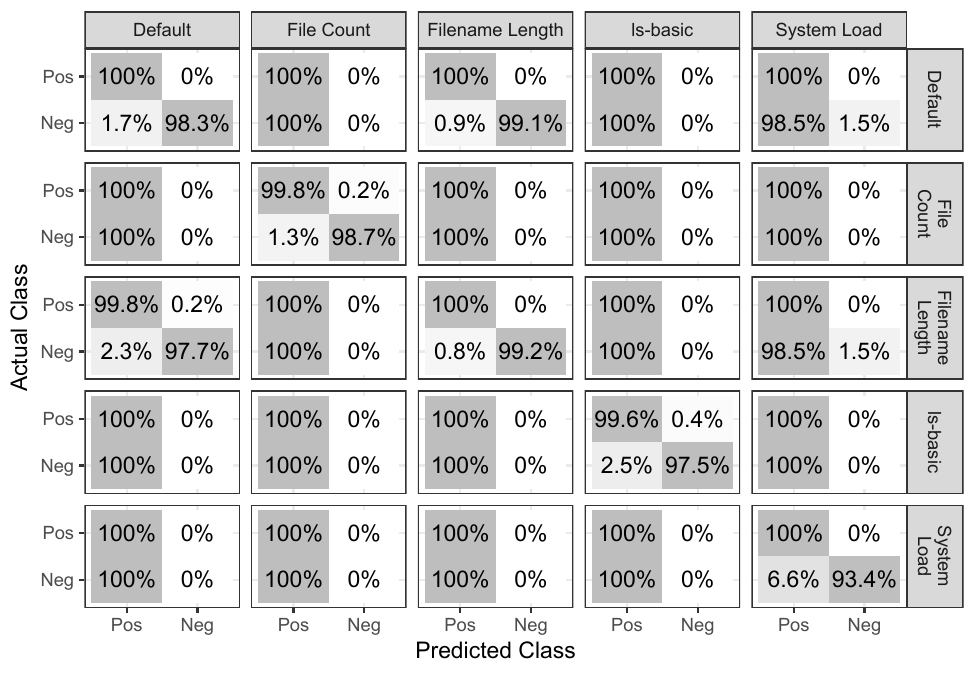}
	\caption{Confusion matrix for function-grouping.}
	\label{fig:confusion_offline_fun}
	\vspace{-7pt}
\end{figure}

\begin{figure}
\centering
\includegraphics[width=\columnwidth]{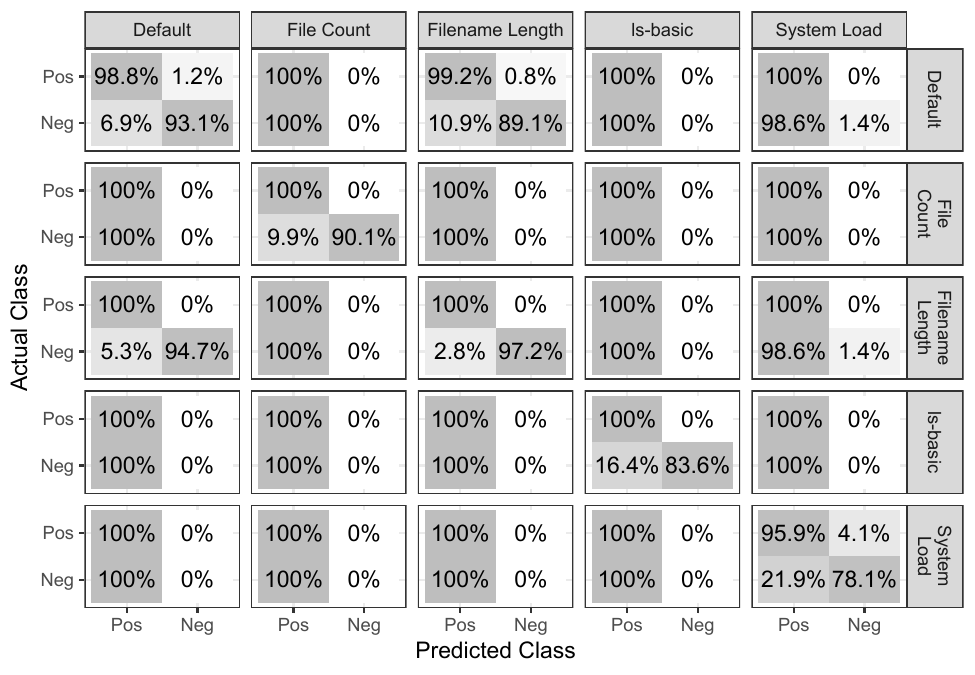}
\caption{Confusion matrix for sequence-grouping.}
\label{fig:confusion_offline_seq}
\vspace{-7pt}
\end{figure}

We fine-tune the detection threshold $\theta$ to maximize the F1 score, but notice that most $p$-values of rootkit batches are truncated to $0$, meaning that best results are achieved when the threshold is set to a very small but non-zero value, such as $10^{-10}$. Figure \ref{fig:confusion_offline_fun} shows the confusion matrix for our detection results, where we state the relative number of batches that end up in each cell. Note that due to the fact that each scenario is evaluated separately and the prediction only differentiates between the normal and rootkit class but not the scenario, it is necessary to consider the values for each scenario in the columns on its own. For example, the top left corner of the matrix shows that all rootkit batches and 98.4\% of all normal batches from the default scenario have been correctly detected as such, while 1.7\% of the normal batches have been incorrectly detected as anomalous and thus contribute to the false positive counts. For all other blocks in that column only the testing data has been changed, e.g., the block directly below shows that all normal and rootkit batches of the file count scenario are detected as anomalous. We leave these values in the confusion matrix for information, but emphasize that they do not contribute to the metric counts. Interestingly, the default scenario seems to be suitable as training data for the filename length scenario and vice versa. This confirms that filename lengths have virtually no impact on delta times, which was already indicated by the right plot in Fig. \ref{fig:pca}. Overall, function-grouping seems to be suitable to correctly classify most of the batches. In comparison to that, the confusion matrix in Fig. \ref{fig:confusion_offline_seq} suggests that sequence-grouping is significantly more prone to false positives across all scenarios.

\begin{figure}
\centering
\includegraphics[width=\columnwidth]{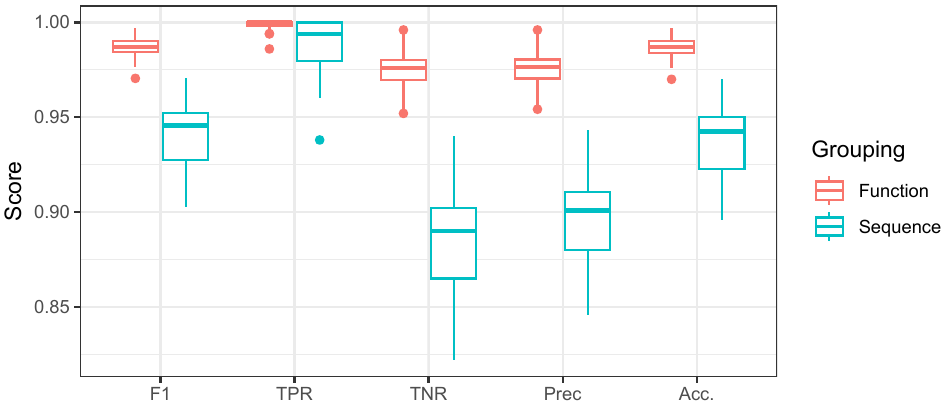}
\caption{Comparison of evaluation results.}
\label{fig:metrics_offline}
\vspace{-7pt}
\end{figure}

Figure \ref{fig:metrics_offline} compares the evaluation metrics for function- and sequence-based grouping and confirms that the latter is unable to keep up with the performance of the former. In particular, the true negative rate ($TNR$) of sequence-grouping, which is driven by false positives, is comparatively low. Moreover, function-grouping correctly detects most of the rootkit batches and yields almost perfect true positive rates ($TPR$). Overall, both strategies perform reasonably well on the data, with median F1 scores of $98.7\%$ and $94.6\%$ achieved by function- and sequence-grouping respectively.

\subsubsection{Online Detection} \label{online}

In realistic settings it is usually non-trivial to ensure that training sets are free of anomalies, because manual and forensic investigation of data for rootkit traces (or the absence thereof) is a time-intensive task that requires specific domain knowledge. Moreover, it is usually not possible to assume that batches of the same class and scenario are identically distributed over time, since real systems are affected by concept drift and more recently collected batches usually represent the current system behavior better than ones that have been collected some time ago. Accordingly, while the methodology presented in the previous section is suitable to measure and compare detection capabilities for evaluation, real applications often require online detection that enables incremental processing of batches.

In addition to offline detection, i.e., random selection of training data from normal batches as described in the previous section, we conduct an experiment for online detection where data is processed chronologically and only the most recent batches are used for training. To this end we run a sliding window of size $w=50$ over the chronologically sorted batches and use them to predict whether the batch following just after that sequence of batches corresponds to normal or rootkit behavior. We thereby leave the order of the batches unchanged from the sorting displayed in Table \ref{tab:datasets}, meaning that batches from each scenario are processed one after another and normal batches of each scenario are processed first before switching to rootkit batches of the same scenario. 

\begin{figure*}
	\centering
	\includegraphics[width=\textwidth]{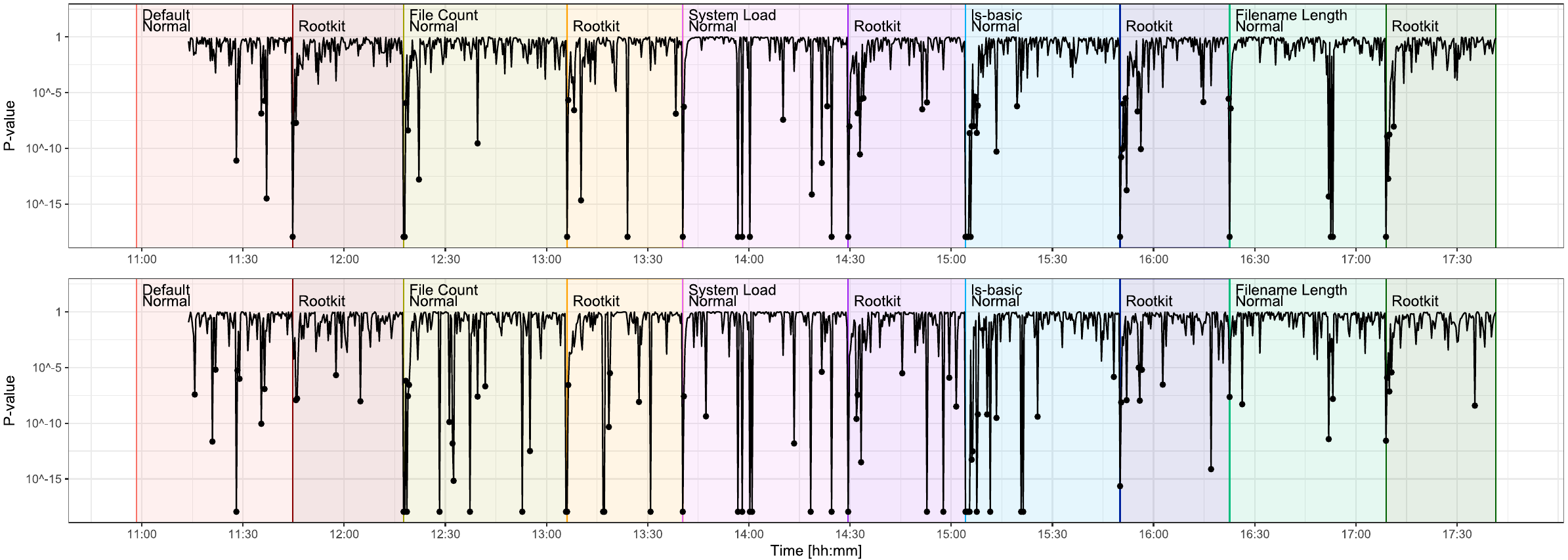}
	\caption{Online detection of function-grouping (top) and using sequence-grouping (bottom).}
	\label{fig:online}
	\vspace{-7pt}
\end{figure*}

Figure \ref{fig:online} visualizes the $p$-values computed with the sliding window method for probes iterate\_dir-enter:iterate\_dir-return using function-grouping (top) and probes filldir64-return:filldir64-enter using sequence-grouping (bottom). In both plots, batches occurring after the switches from normal to rootkit batches and vice versa receive low $p$-values since their delta values do not fit the distributions of delta values of the preceding batches. Similar to the observations made in Sect. \ref{offline}, sequence-grouping appears to suffer from more false positives than function-grouping.

To numerically assess the detection performance we again need to count correct and incorrect classifications. However, online detection with sliding windows implies that training data comprises both normal and rootkit batches when the window passes over the point where these two classes meet. To account for the fact that the model is unreliable in that case, we compute the evaluation metrics as follows. Positives are the first normal and rootkit batches occurring just after the switching from one class to another. We count them as true positives ($TP$) if they are detected as anomalous and false negatives ($FN$) otherwise. Negatives are all batches that occur after at least $w$ batches from the same class have been processed to ensure that the predictions are only counted when the model is trained with either normal or rootkit batches, but not a mix thereof. We count them as false positives ($FP$) if they are detected as anomalous and true negatives ($TN$) otherwise. Combining the $p$-values of all probes and computing the evaluation metrics using the equations stated in Sect. \ref{offline} we obtain $TPR = 100\%$, $TNR = 97\%$, $Prec = 28.1\%$, $Acc = 97\%$, and $F1 = 43.9\%$ for function-grouping and $TPR = 100\%$, $TNR = 90.7\%$, $Prec = 11.3\%$, $Acc = 90.8\%$, and $F1 = 20.2\%$ for sequence-grouping. This confirms that function-grouping outperforms sequence-grouping across all metrics, which aligns with the results obtained from offline evaluation.

\section{Discussion} \label{discussion}

This paper demonstrates the detectability of rootkits through measurement of time delays of function calls within system calls. Thereby, these delays are caused by additional code that needs to be executed when rootkits wrap around certain functions to hide their presence and other objects, e.g., check and skip certain filenames when users enumerate the contents of directories. Based on our insights gained from this work, we answer the research questions as follows. \textit{RQ1: What system calls enable the observation of rootkits that hide files?} Based on a review of existing open-source rootkits and specifically their methods to hide files, we found that among many system calls that may be affected by rootkits, the getdents system call is the most relevant as it is the main interface to list file contents on Linux. Thereby, we point out that rootkits do not necessarily have to target the entire system call but may also wrap around some of its inner functions, such as filldir. \textit{RQ2: How can delays of relevant function calls be observed?} Absolute time measurements can be collected through injection of eBPF probes that attach to enter and return points of functions within the kernel. Based on these measurements, delta times between any combination of probes can be computed. For example, this allows to compute the time to execute entire functions (function-grouping) or the time between two subsequently invoked functions (sequence-grouping). \textit{RQ3: To what degree can anomaly detection techniques leverage system call function timings to uncover hidden rootkit activities?} Our proof-of-concept anomaly detection approach relies on statistical tests at certain quantiles of delta time distributions to recognize time shifts in comparison to normal behavior models. The results of our evaluation suggest that this approach is able to detect rootkit activity with high accuracy; however, false positives have been noticed as a problem that could limit practical applicability. In particular, system conditions have a significant influence on the delta times, which could trigger many false positives in highly dynamic systems.

We foresee several ways to address the issue of negative influence of varying system conditions on the detection accuracy. First, practical implementations of our detection method could make use of dynamic probing intervals rather than regular intervals of $10$ seconds used in our data generation procedure. In particular, the probing mechanism could wait for the system to be in an idle state to reduce noise that affects execution times of functions and interferes with time measurements. In addition, detected anomalies could trigger additional probing in short intervals that allow to determine whether the observed shift of delta times is constant and persists over time as it should be expected when a rootkit permanently wraps a function, or whether the currently processed batch of time measurements should be considered an outlier and the system can be regarded as normal despite an anomaly. Such an approach could also be used to assign confidence scores to predictions. Second, the time measurements themselves could be made more robust by assigning high priorities to the probing mechanisms so that other concurrent system processes do not interfere with the observed function call timings. Third, the detection approach based on statistical testing could be replaced with more generic alternatives. In particular, neural networks are well suited to ingest the complex and non-linear nature of delta time distributions and could thus be used to assign anomaly scores to system states. Fourth, in comparison to our approach that uses only data from a single scenario for training, other approaches could leverage data from multiple scenarios for training to generate a single normal behavior model that enables classification of batches from any of these scenarios. As visible in the right plot of Fig. \ref{fig:pca}, delta times are sufficiently different across scenarios so that batches can be first assigned a scenario through clustering with sequence-grouped delta times before carrying out detection with function-grouped delta times. Thereby, our procedure of generating data sets could be extended with new scenarios that introduce other forms of noise to obtain data sets with even more variation of normal behavior. Alternatively, it is also possible to mix two or more existing scenarios so that multiple sources of noise occur at the same time. We refer to the work by Singh et al. \cite{singh2017detection}, who introduce noise by interacting with various programs, such as browsers and benchmark tools. Fifth, in contrast to our semi-supervised approach, supervised approaches could either make use of labeled instances of specific scenarios or even batches generated when rootkits are active to further improve classification and detection performance. 

Several modifications could be made to our rootkit to extend the evaluation. In addition to wrapping filldir, the rootkit can alternatively wrap the entire getdents system call for file hiding. It could thus be interesting to investigate to what degree the detection of these methods differ when it comes to detection. Moreover, the rootkit does not only support file hiding, but also process hiding. After experimenting with both functionalities, we noticed that there is no significant difference when it comes to detection, which is why we focused on the more simple case of file hiding in this paper. Finally, while the implementation of our approach is designed for and evaluated on Linux machines, the concept of system calls is agnostic to operating systems. Therefore, our concept of measuring low-level function timings for rootkit detection may be transferred to other operating systems, such as Windows \cite{bravo2011proactive, singh2017detection}.

\section{Conclusion} \label{conclusion}

This paper presents a semi-supervised and anomaly-based approach that leverages statistical testing for rootkit detection based on kernel function timings. The main idea behind this concept is that rootkits need to inject code that modifies the outcome of specific kernel functions to hide their presence from users or detection programs, which increases the runtime of these functions. To facilitate measurement of these time intervals, we present a framework that allows injects probes into the kernel, attaches them to enter and return points of functions, and polls time stamps for invocations of selected functions. Thereby, we found that the getdents system call and its inner functions are of particular relevance as they are key to enable hiding capabilities of rootkits. We convert absolute time measurements into delta times using two strategies that focus on functions and sequential steps of program workflows respectively. Our analysis suggests that function-grouping has advantages when it comes to the detection of shifted delta times, while sequence-grouping is superior when it comes to classification of system states. We collect batches of delta times in five different scenarios using a custom rootkit and test our detection approach in offline and online settings. The results of our evaluation indicate high detection accuracy and leaves many interesting research opportunities for future work, such as mechanisms for dynamic probing, rootkit detection across different system states, and experiments with machine learning models other than statistical tests.

\begin{acks}
Parts of this work were carried out in course of a Master’s Thesis at the Vienna University of Technology \cite{alton2024root}. The work in this paper has received funding from the European Union - Horizon Europe Research and Innovation programme under GA no. 101168144 (MIRANDA) and European Defence Fund under GA no. 101121403 (NEWSROOM). Views and opinions expressed are however those of the author(s) only and do not necessarily reflect those of the European Union. The European Union cannot be held responsible for them.
\end{acks}

\bibliographystyle{ACM-Reference-Format}
\bibliography{sample-sigconf}


\end{document}